\documentclass[12pt,a4paper]{article}
\pdfoutput=1
\usepackage{jheppub}
\usepackage[sort&compress]{natbib}
\usepackage{url}
\usepackage{hyperref}
\usepackage{amsfonts}
\usepackage{epsfig}\usepackage{dcolumn}
\usepackage{amssymb}
\usepackage{bm}
\usepackage{slashed}
\graphicspath{{./figuras/}}
\usepackage{subfigure}
\usepackage{graphicx}

\title{Melanopogenesis: Dark Matter of (almost) any Mass and Baryonic Matter from the Evaporation of Primordial Black Holes weighing a Ton (or less)}

\author{Logan Morrison,}
\author{Stefano~Profumo, and}
\author{Yan Yu}

\affiliation{Department of Physics and Santa Cruz Institute for Particle Physics,\\
University of California, Santa Cruz, CA 95064, USA}

\emailAdd{loanmorr@ucsc.edu}
\emailAdd{profumo@ucsc.edu}
\emailAdd{yanyu@ucsc.edu}

\abstract{The evaporation of primordial black holes with a mass in the $1\  {\rm gram}\lesssim M_{\rm PBH}\lesssim$1000 kg range can lead to the production of dark matter particles of almost any mass in the range $0.1\ {\rm MeV}\lesssim m_{\rm DM}\lesssim 10^{18}$ GeV with the right relic density at very early times, $\tau\lesssim 10^{-10}$ s. We calculate, as a function of the primordial black holes mass and initial abundance, the combination of dark matter particle masses and number of effective dark degrees of freedom leading to the right abundance of dark matter today, whether or not evaporation stops around the Planck scale. In addition, since black hole evaporation can also lead to the production of a baryon asymmetry, we calculate where dark matter production and baryogenesis can concurrently happen, under a variety of assumptions: baryogenesis via grand unification boson decay, via leptogenesis, or via asymmetric co-genesis of dark matter and ordinary matter. Finally, we comment on possible ways to test this scenario.}

\begin{document} 
\maketitle
\flushbottom

\section{Introduction}

The nature of non-baryonic dark matter and the origin of the matter-antimatter asymmetry in the universe are two of the most pressing questions at the junction of particle physics and cosmology (see e.g. \cite{Profumo:2017hqp, Balazs:2014eba}).  The direct detection of gravitational waves \cite{Abbott:2016blz} has triggered renewed interest in the role black holes might play in connection with these two outstanding open questions (see e.g. \cite{Bird:2016dcv, Hamada:2016jnq, Cline:2018fuq} and references therein).

The idea that black hole evaporation can lead to the production of particles in the early universe and possibly to the generation of a baryon asymmetry or of dark matter has a rather long history. In fact, it dates back to early work by Hawking \cite{1975CMaPh..43..199H} and Zel'dovich \cite{1976JETPL..24...25Z}, and to subsequent work by Carr \cite{1976ApJ...206....8C}, where it was pointed out that evaporation might produce a baryon asymmetry because of intrinsic $CP$-violating effects, or because of accidental statistical excesses. Subsequent work invoked GUT-scale physics, which generically violates $CP$ and baryon number, as a culprit for the generation of the matter-antimatter asymmetry, and specifically $CP$- and $B$-violating decays of GUT-scale particles produced in PBH evaporation. This was  first (to our knowledge) envisioned in Ref.~\cite{Toussaint:1978br}, and further elaborated upon in Ref.~\cite{Turner:1979bt,Grillo:1980rt,Hawking:1982ga}.

More recently, studies have considered the possibility that primordial black holes be responsible for both the generation of a baryon asymmetry {\em and} of the dark matter. Specifically, Ref.~\cite{Baumann:2007yr} considers the concurrent generation of a baryon asymmetry from GUT boson decays in inflationary and in ekpyrotic/cyclic models, and the possible presence of dark matter in the form of Planck-scale relics from the evaporation of light primordial holes. The key assumption in Ref.~\cite{Baumann:2007yr} is that the black holes dominate the universe's energy budget by the time they decay. More recently, Ref.~\cite{Fujita:2014hha} dealt with the possibility of a generation of the matter asymmetry from asymmetric sterile neutrino decays, through leptogenesis, and the co-genesis of dark matter from Hawking evaporation. Here, the assumption is, again, that the energy density of the universe is dominated by the primordial black holes before they evaporate. Ref.~\cite{Fujita:2014hha} additionally allows for a possible entropy injection episode {\em after} the evaporation of the primordial holes. Ref.~\cite{Lennon:2017tqq} and \cite{Allahverdi:2017sks} recently focused on production of dark matter from black hole evaporation (the second specifically WIMP dark matter), while Ref.~\cite{Bell:1998jk} and \cite{Hook:2014mla} considered scenarios technically similar to asymmetric dark matter production from black hole evaporation that we also entertain below, although the first study in the context of mirror matter and the second in the context of asymmetric Hawking radiation.

In the present study, we are not concerned with {\em how} primordial black holes were produced following inflation (any pre-existing population would be presumably inflated away). The conditions under which a cosmologically relevant population of primordial black holes are produced have been known for a long time, ever since the seminal work of Zel'dovich and Novikov from over a half century ago \cite{1966AZh....43..758Z}. We will not review them here, referring the Reader to the extensive existing literature \citep[see e.g.][]{Hawking:1971ei,1978AZh....55..216N,Shibata:1999zs,Musco:2012au,Harada:2013epa,Nakama:2013ica}. In fact, in what follows we will use the relative abundance of primordial black holes at formation as a free parameter.

We intend to study here the generic possibility that primordial black hole evaporation at very early times (much earlier than, say, the epoch when the synthesis of light elements occurs) plays a key role in the genesis of dark matter and/or the baryon asymmetry in the universe. We will be agnostic about the fate of black holes as their mass approaches the Planck scale, and consider both complete evaporation, and the possibility that evaporation stops around or below that scale, leaving Planck-scale relics. We  consider a variety of scenarios for the genesis of the matter-antimatter asymmetry, including GUT-scale baryogenesis, leptogenesis, and asymmetric co-genesis of dark and ordinary matter. Finally, we discuss possible tell-tale signals of this general framework.

The structure of this manuscript is as follows: in the following section \ref{sec:framework} we outline our framework; sec.~\ref{sec:dm} and \ref{sec:bau}, respectively, discuss the generation of dark matter and of a baryon asymmetry from primordial black hole evaporation, after a general discussion of particle production from hole evaporation; sec.~\ref{sec:results} presents our results and a discussion thereof, and the final sec.~\ref{sec:conclusions} concludes.

\section{Outline of the framework}\label{sec:framework}

In this study, we explore the possibility that primordial black hole evaporation produced both the dark matter and the baryon asymmetry. We entertain the possibility that the dark matter be a species $\chi$, with mass $m_{\chi}$ belonging to a {\em dark sector} with a number of degrees of freedom $g_\chi$ that ranges, for definiteness, from 1 to 100. We do not make any assumption on the details of the dark sector spectrum. In principle, these details affect the temperature dependence of the number of degrees of freedom in the early universe; however, as we show below, the quantities relevant for us all depend quite weakly on that. We therefore neglect details of the dark sector spectrum and its impact on the number of degrees of freedom as a function of temperature, and assume the dark sector is all degenerate at the same mass scale $m_{\chi}$.

We assume that the $\chi$ never attains thermal equilibrium after being produced (unless we specify otherwise, as is the case when we discuss the case of asymmetric dark matter), and that no processes exist that can {\em freeze in} any significant abundance of $\chi$ after black hole evaporation \cite{freezein}. We also entertain both the possibility that black hole evaporation stops at a black hole mass $f M_{\rm Pl}$, with $M_{\rm Pl}$ the reduced Planck mass, leading to a multi-component dark matter scenario consisting of the species $\chi$ produced by evaporation and of the Planck-scale relics, and the second possibility that evaporation leads to the complete disappearance of the primordial holes ($f=0$) and thus to a single-dark matter scenario.

As far as the production of the baryon asymmetry, we entertain three distinct possibilities:
\begin{enumerate}
    \item That the baryon asymmetry is produced by the $B$-, $L$-, and $CP$-violating decays of a GUT boson $X$ of mass $10^{15}\lesssim m_X/{\rm GeV}\lesssim 10^{17}$;
    \item That the baryon asymmetry stems from leptogenesis induced by the $CP$ violating decays of heavy right-handed neutrinos with masses at the scale $M_\nu$;
    \item That both the baryon asymmetry and the dark matter are produced by the decay of heavy right-handed neutrinos (or some other massive species) into both dark matter/dark sector fields and standard model/visible sector fields, with the dark matter produced by evaporation annihilating away.
\end{enumerate}

Unlike in previous studies, here we let the initial abundance $\beta$ of primordial black holes at the epoch of formation, at a time $t_i$ normalized to the radiation density,
\begin{equation}
    \beta\equiv\frac{\rho_{\rm PBH}(t_i)}{\rho_{\rm rad}(t_i)},
\end{equation}
to be a free parameter. 

As mentioned above, we do consider the possibility that relics of mass $f M_{\rm Pl}$ are left over from PBH evaporation. In this case the approximate constraint is \cite{Carr:1994ar}
\begin{equation}
    \beta(M_{\rm PBH})\lesssim 2.8\times 10^{-28}\ f^{-1}\left(\frac{M_{\rm PBH}}{M_{\rm Pl}}\right)^{3/2}.
\end{equation}

We assume for simplicity that the mass function of primordial black holes (PBH) is monochromatic and centered at a mass $M_{\rm PBH}$ (see e.g. \cite{1978AZh....55..216N} for a motivation to this assumption, and \cite{Lehmann:2018ejc} for a recent study of the optimal mass function for dark matter in the form of PBH).  The range of viable black hole masses is constrained from below by the requirement that the black holes form when the Hubble rate is at or below the Hubble rate during inflation $H_*$. The latter is constrained by Planck observations \cite{Akrami:2018odb} to be
\begin{equation}
    \frac{H_*}{M_{\rm Pl}}<2.7\times 10^{-5}\quad ({\rm 95\%\ C.L.}). 
\end{equation}
If, as we assume here, primordial black holes form during the radiation domination epoch, their initial mass is
\begin{equation}\label{eq:mpbh}
    M_{\rm PBH}=\gamma\frac{4\pi}{3}\rho H^{-3},
\end{equation}
where $\gamma\sim0.2$ \cite{Carr:2009jm}, and the Planck limit translates to
\begin{equation}\label{eq:hubbleconst}
    \frac{M_{\rm PBH}}{M_{\rm Pl}}>\frac{4\pi\gamma}{2.7\times 10^{-5}}\simeq 9.1\times 10^4\quad ({\rm 95\%\ C.L.}).
\end{equation}

Notice that the density of PBHs depends on the spectrum of primordial density perturbations $\delta\equiv\delta\rho/\rho$, which at a given epoch has a certain probability distribution $P(\delta)$, for instance of Gaussian form; the fraction of energy density of the universe that collapses to PBHs is then
$$
\beta(M_{\rm PBH})=\int_{\delta_c}^\infty P(\delta)d\delta,
$$
with $\delta_c\sim\gamma$ the critical overdensity; in the case of Gaussian density perturbations with mass variance $\sigma$ at horizon crossing, $\beta\sim\exp(-\delta_c^2/(2\sigma^2)$; since generally $\delta_c\gg\sigma$, only a small fraction of the energy density of the universe thus collapses to form PBHs, explaining why $\beta\ll 1$.

The upper limit to the mass of the PBH, in our framework, derives from the requirement that evaporation happens well before Big Bang Nucleosynthesis. In principle this is not a hard requirement, but, should it not be satisfied, the resulting limits on $\beta(M_{\rm PBH})$ would rule out any significant production of either dark matter or the baryon asymmetry from PBH evaporation, defeating the point of our study.

\begin{table}[]
    \centering
    
    \begin{tabular}{|c|c|c|c|}
    \hline
    Mass (g)     & $T_H$ (GeV) & $\tau$ (s) & $T_{\rm evap}=T(\tau)$ (GeV) \\
    \hline
    $5M_P\simeq10^{-4}$ & $1.7\times 10^{17}$ & $10^{-41}$ & $2\times 10^{17}$ \\
    $1$ & $1.7\times 10^{13}$ & $4\times10^{-29}$ & $2\times 10^{11}$ \\
    $10^3$ & $1.7\times 10^{10}$ & $4\times10^{-20}$ & $6\times 10^{6}$ \\
    $10^6$ & $1.7\times 10^{7}$ & $4\times10^{-11}$ & 200 \\
    $10^9$ & $1.7\times 10^{4}$ & $0.04$ & 0.006 \\
    $10^{12}$ & $17$ & $4\times10^{7}\sim1$ yr & $\sim1$ keV \\
            \hline
    \end{tabular}
    \caption{Mass, Hawking-Gibbons temperature, lifetime, and temperature corresponding to the evaporation time for a few illustrative black hole masses.}
    \label{tab:pbh}
\end{table}

We now quickly summarize PBH evaporation: upon integrating the mass loss rate of PBH from Hawking-Gibbons evaporation from a black body at temperature $T_H\equiv M_{\rm Pl}^2/M_{\rm PBH}$, and neglecting grey-body factors \cite{Page:1976df}, the lifetime of a PBH of mass $M_{\rm PBH}$ is
\begin{equation}
    \tau=\frac{160}{\pi g}\frac{M_{\rm PBH}^3}{M_{\rm Pl}^4},
\end{equation}
where $g$ are the number of degrees of freedom of the radiated particles (including the 7/8 factor for fermions), which, here, are all particles with mass smaller than $T_H$. The radiation bath temperature corresponding to the time $\tau$ is calculated in the standard way for a radiation-dominated cosmology, assuming instantaneous thermalization of the evaporation products: using Friedmann's equation \cite{Fujita:2014hha}
\begin{equation}\label{eq:evap}
    \frac{\pi^2}{30}g_* T_{\rm evap}^4=3M_{\rm Pl}^2 H^2_{\rm evap}\simeq 3M_{\rm Pl}^2\tau^{-2}
\end{equation}
with $g_*$ the number of effective relativistic degrees of freedom, we get
\begin{equation}\label{eq:tevap}
    \frac{T_{\rm evap}}{M_{\rm Pl}}\simeq 0.77\left(\frac{g_*}{100}\right)^{-1/4}\left(\frac{g}{100}\right)^{1/2}\left(\frac{M_{\rm Pl}}{M_{\rm PBH}}\right)^{3/2},
\end{equation}
which is of course only valid if evaporation ends during radiation domination, which is always the case for us.

We list in Table \ref{tab:pbh} masses, Hawking-Gibbons temperatures, lifetimes, and temperature corresponding to the evaporation time for a few illustrative black hole masses. The table shows that in order to avoid impacting the synthesis of light elements (at times of $t\sim{\mathcal O}(1\ {\rm sec})$, the PBH under consideration here are required to be lighter than approximately 1,000 t, i.e. $10^9$ grams.

\section{Particle production from Primordial Black Hole Evaporation}
We intend to calculate the abundance of right-handed neutrinos or dark matter (generically, of any massive particle $X$) produced by the evaporation of PBH in an adiabatically expanding universe, if the relative density of PBH to radiation at PBH formation time $t_i$ is $\beta(t_i)=\rho_{\rm PBH}/\rho_{\rm rad}$. Indicating with $N_X$ the number of particles $X$ produced in the evaporation of one single hole, the number-to-entropy density of particles $X$ at the present epoch is
\begin{equation}\label{eq:Xabund}
    \frac{n_X}{s}(t_{\rm now})=N_X\frac{n_{\rm PBH}}{s}(t_i)=N_X Y_i.
\end{equation}
To calculate $Y_i$, we use the definition of $\beta$,
\begin{equation}\label{eq:betaeq}
    \beta=M_{\rm PBH}\frac{n_{\rm PBH}(t_i)}{\rho_{\rm rad}(t_i)}=M_{\rm PBH}\frac{s(t_i)}{\rho_{\rm rad}(t_i)}Y_i=\frac{4}{3}\frac{M_{\rm PBH}}{T_i}Y_i
\end{equation}
where in the last equality we have assumed that at very large temperatures $g\simeq g_*$, the latter indicating the entropic degrees of freedom. From Eq.~(\ref{eq:mpbh}) we then get
\begin{equation}\label{eq:Tinitial}
    \frac{T_i}{M_{\rm Pl}}\simeq 0.87\left(\frac{M_{\rm Pl}}{M_{\rm PBH}}\right)^{1/2}\left(\frac{g_*}{100}\right)^{-1/4}.
\end{equation}
Now, substituting Eq.~(\ref{eq:Tinitial}) into Eq.~(\ref{eq:betaeq}) and the expression for $Y_i$,
\begin{equation}
    Y_i=0.65\ \beta\  \left(\frac{g_*}{100}\right)^{-1/4}\left(\frac{M_{\rm Pl}}{M_{\rm PBH}}\right)^{3/2},
    \label{eq:yi}
\end{equation}
into Eq.~(\ref{eq:Xabund}), we get
\begin{equation}
    \frac{n_X}{s}(t_{\rm now})\ \simeq\ 0.65\ \beta\  N_X\left(\frac{g_*}{100}\right)^{-1/4}\left(\frac{M_{\rm Pl}}{M_{\rm PBH}}\right)^{3/2}.
\end{equation}
The calculation of $N_X$ follows Ref.~\cite{Baumann:2007yr}. Assume that $M_X<T_H=M_{\rm Pl}^2/M_{\rm PBH}$, and assume that evaporation does not stop at $M_{\rm Pl}$, i.e. $f=0$. In this case, 
\begin{equation}\label{eq:NX}
    N_X\simeq\frac{g_X}{g}\int_{M_{\rm PBH}}^0\frac{-dM}{3T}=\frac{g_X}{g}\int_{T_H}^\infty\frac{M_{\rm Pl}^2}{3T^3}dT=\frac{g_X}{6g}\left(\frac{M_{\rm PBH}}{M_{\rm Pl}}\right)^2.
\end{equation}
In the first equality, we assumed that the number of radiated particle is given by the ratio of the radiation energy from PBH evaporation, $-dM$, divided by the mean energy of a black-body of temperature $T$, $\langle E \rangle_T\simeq 3T$ (we notice that this is an approximate result that assumes the particles produced at evaporation to be spin zero, as well as a trivial, constant absorption cross section; in more realistic setups, there is a complicated dependence of the factor in front of $T$ on the spectrum and spin of the particles the PBH evaporates to, see e.g. the classical literature on this point, Ref.~\cite{Page:1976df,Page:1976ki,Page:1977um}). 

The ratio of the degrees of freedom $g_X/g$, where $g=g_X+g_{\rm SM}$, the latter $g_{\rm SM}$ indicating the ``standard model'' degrees of freedom, corresponds to the approximate ratio of emitted $X$ particles, neglecting the effects of charge and spin on the evaporation rate, and in the second equality we used the relation between Hawking-Gibbons temperature and black hole mass. In the case where evaporation stops when the black hole mass is equal to $fM_{\rm Pl}$, the equation above is modified as follows:
\begin{equation}
    N_X^f\simeq\frac{g_X}{6g}\left(\left(\frac{M_{\rm PBH}}{M_{\rm Pl}}\right)^2-f^2\right),
\end{equation}
and of course reduces to the result above if $M_{\rm PBH}\gg f M_{\rm Pl}$; given the constraint in Eq.~(\ref{eq:hubbleconst}) above, unless $f\gtrsim 10^4$, a range theoretically unmotivated for evaporation to stop, Eq.~(\ref{eq:NX}) above is perfectly adequate, and we shall use it from now on.

Notice that in principle massive particles $X$ can be produced by PBH evaporation even if the {\em initial} Hawking-Gibbons temperature is {\em lower} than $m_X$. In that case, $X$ production proceeds from the moment when $t_H=M_X$, thus
\begin{equation}
    N_X\simeq\frac{g_X}{g}\int_{M_X}^{\infty}\frac{M_{\rm Pl}^2}{3T^3}dT=\frac{g_X}{6g}\left(\frac{M_X}{M_{\rm Pl}}\right)^{-2}.
\end{equation}
In summary, given a sector with $g_X$ degrees of freedom and a stable relic $X$ of mass $M_X$, the cosmological abundance $\Omega_X=\rho_X/\rho_c$, with $\rho_c$ the critical energy density of the universe, is
\begin{eqnarray}\label{eq:dmabundance}
    \Omega_X&=&\frac{M_X}{\rho_c}\frac{n_X}{s}(t_{\rm now})s_{\rm now}=\frac{M_X}{\rho_c}N_XY_i s_{\rm now}\\
    \nonumber &\simeq&0.11\ \beta\ \frac{M_X s_{\rm now}}{\rho_c} \left(\frac{g_*}{100}\right)^{-1/4}\left(\frac{g_X}{g}\right)\left(\frac{M_{\rm PBH}}{M_{\rm Pl}}\right)^{1/2},\qquad M_X<M_{\rm Pl}^2/M_{\rm PBH},\\
    \nonumber &\simeq&0.11\ \beta\ \frac{M_X s_{\rm now}}{\rho_c} \left(\frac{g_*}{100}\right)^{-1/4}\left(\frac{g_X}{g}\right)\left(\frac{M_{\rm Pl}^7}{M_{\rm PBH}^3 M_X^4}\right)^{1/2},\qquad M_X>M_{\rm Pl}^2/M_{\rm PBH}.
\end{eqnarray}
Since \cite{Aghanim:2018eyx}
\begin{equation}
    \rho_c=1.0537\times 10^{-5}\ h^2\ \frac{\rm GeV}{{\rm cm}^3}\simeq4.78\times 10^{-6}\left(\frac{h}{67.37}\right)^2\ \frac{\rm GeV}{{\rm cm}^3}
\end{equation}
and \cite{Aghanim:2018eyx}
\begin{equation}
    s_{\rm now}=2,891.2\left(\frac{T_{\rm CMB}}{2.7255}\right)^2\ {\rm cm}^{-3},
\end{equation}
we can recast the equations above, for $g_*=106.75+1$ (which assumes $g_{\rm DM}=1$ for definiteness), as
\begin{eqnarray}\label{eq:dmabundance2}
    \Omega_X&\simeq&6.5\times 10^7\ \beta\ \left(\frac{M_X}{\rm GeV}\right) \left(\frac{g_X}{g}\right)\left(\frac{M_{\rm PBH}}{M_{\rm Pl}}\right)^{1/2},\qquad M_X<M_{\rm Pl}^2/M_{\rm PBH},\\
    \nonumber &\simeq&6.5\times 10^7\ \beta\ \left(\frac{M_X}{\rm GeV}\right)\left(\frac{g_X}{g}\right)\left(\frac{M_{\rm Pl}^7}{M_{\rm PBH}^3 M_X^4}\right)^{1/2},\qquad M_X>M_{\rm Pl}^2/M_{\rm PBH}.
\end{eqnarray}

Since the radiation energy density redshifts like $a^{-4}$, with $a$ the scale factor, while pressure-less matter redshifts as $a^{-3}$, for sufficiently large $\beta$, the universe could become matter-dominated by PBH prior to evaporation. This condition can be expressed as 
\begin{equation}
    \beta_f=\frac{\rho_{\rm PBH}(T_{\rm evap})}{\rho_{\rm rad}(T_{\rm evap})}>1.
\end{equation}
Assuming entropy conservation, we can relate the initial value of $\beta$ to $\beta_f$ in the equation above,
\begin{equation}\label{eq:betaf}
    \beta_f=\beta\frac{T_i}{T_{\rm evap}}\simeq 1.1\left(\frac{g}{100}\right)^{-1/2}\left(\frac{M_{\rm PBH}}{M_{\rm Pl}}\right),
\end{equation}
where we used Eq.~(\ref{eq:Tinitial}) and (\ref{eq:tevap}) to express $T_i$ and $T_{\rm evap}$ as functions of $M_{\rm PBH}$. If PBH get to dominate the energy density of the universe prior to evaporation, the number density of particles produced by evaporation ceases to depend on $\beta$, as pointed out by Ref.~\cite{Fujita:2014hha}, as a result of the balance between the dilution of the number density of particles produced by evaporation and of the additional particles resulting from $\beta_f>1$. In all our results, we highlight the parameter space on the $(M_{\rm PBH},\beta)$ plane where $\beta_f=1$ (for $g_\chi=1$ plus Standard Model degrees of freedom) with a thick purple line; for $\beta$ values above that line, our results are $\beta$-independent.  

\subsection{Dark Matter from PBH evaporation}\label{sec:dm}

We consider three mechanisms for dark matter production from PBH evaporation: direct production from evaporation, Planck-scale relics from evaporation, and asymmetric dark matter production. We postpone the discussion of the latter to the next section, and Eq.(\ref{eq:dmabundance}) directly gives the abundance of dark matter from PBH evaporation. 

If evaporation stops at a mass scale $fM_{\rm Pl}$, the cosmological abundance of Planck-scale relics, $\Omega_P=(f M_{\rm Pl}n_{\rm PBH}(t_{\rm now}))/\rho_c$ is given by
\begin{eqnarray}
    \nonumber \Omega_P&=&\frac{f M_{\rm Pl}}{\rho_c}Y_i s_{\rm now}=0.65\ \beta\ f\ \frac{M_{\rm Pl}s_{\rm now}}{\rho_c}\left(\frac{g_*}{100}\right)^{-1/4}\left(\frac{M_{\rm Pl}}{M_{\rm PBH}}\right)^{3/2},\\
    &\simeq&9.4\times 10^{26} \ \beta\ f\ \left(\frac{M_{\rm Pl}}{M_{\rm PBH}}\right)^{3/2},
\end{eqnarray}
again with $g_*=106.75+1$ in the second equation.

Requiring that the dark matter in the universe have a density $\Omega_{\mathrm{CDM}} \simeq 0.21$ \cite{Aghanim:2018eyx} forces a condition across the model parameters $g_\chi$, $m_{\chi}$, $M_{\rm PBH}$, $f$ and $\beta$ (where we indicated with $\chi$ the dark matter from PBH evaporation) such that
\begin{equation}
    \Omega_{\mathrm{CDM}}=\Omega_\chi(g_\chi,m_{\chi},M_{\rm PBH},\beta)+\Omega_P(f,M_{\rm PBH},\beta).
\end{equation}
Also, wherever $\Omega_P, \Omega_\chi>\Omega_{\rm \mathrm{CDM}}$, the corresponding region of parameter space is excluded as too much dark matter is produced by either evaporation or Planck-scale leftover relics, or both (of course, this assumes no episode of entropy injection that would dilute the relics' density, see e.g. \cite{Fujita:2014hha}). Notice that regions with {\em underabundant} dark matter from either Planck relics or evaporation are not ruled out, since some other dark matter component might provide the remaining part of the observed cosmological dark matter density.

An important constraint on dark matter $\chi$ produced from PBH evaporation comes from the requirement that the dark matter be {\em cold enough} as to avoid disruption of small-scale structures via free-streaming. We follow here the discussion in Ref.~\cite{Fujita:2014hha}: The initial average energy of particles from the evaporation of a hole of mass $M_{\rm PBH}$ is $6T_H=6M_{\rm Pl}^2/M_{\rm PBH}$ (see the derivation below in Eq.~(\ref{eq:aeenergy})).  Because we assume the dark matter particles are never in kinetic or chemical equilibrium, the particle momentum today is simply the redshifted value of the momentum at production,
\begin{equation}\label{eqn:dm_mom_now}
p_{\rm now}=\frac{a_{\rm evap}}{a_{\rm now}}p_{\rm evap},
\end{equation}
where $a_{\rm now}=1$. The energy of the dark matter particle at evaporation can be calculated as follows: the  average energy of particles radiated by a PBH with a Hawking-Gibbons radiation temperature $T_H$ is 3$T_H$ (with the caveats explained above -- this is a simplifying approximation!). The total number of particles emitted by the PBH is approximately
\begin{equation}
    N=\int_0^N dn=\int_{T_H}^\infty\frac{M_{\rm Pl}^2}{3T^3}dT=\frac{1}{6}\left(\frac{M_{\rm Pl}}{T_H}\right)^2.
\end{equation}
The mean energy of the radiated particles is thus 
\begin{equation}\label{eq:aeenergy}
    \bar E\simeq \int_0^N(3T)\frac{dn}{N}=6\left(\frac{T_H}{M_{\rm Pl}}\right)^2\int_{T_H}^\infty\frac{M_{\rm Pl}^2}{T^2}dT=6\left(\frac{T_H}{M_{\rm Pl}}\right)^2\left(\frac{M^2_{\rm Pl}}{T_H}\right)=6T_H.
\end{equation}
Notice that this average energy is different from the average energy at the beginning of evaporation, since it averages over all temperatures from that initial temperature to infinity. Now, since $m_{\chi}<T_H$ in order for the dark matter to be produced, $\bar E\simeq \bar p=p_{\rm evap}$. The last ingredient to calculate the dark matter velocity today is $a_{\rm evap}$. Fixing $a_{\rm now}=1$, the scale factor at matter-radiation equality $a_{\rm eq}=\Omega_r/\Omega_m$. Then, using Friedman's equation, and the fact that  $\rho\sim a^4$ in radiation domination, we have
\begin{equation}\label{eqn:aevap}
    a_{\rm evap}=a_{\rm eq}\left(\frac{\rho_{\rm eq}}{\rho_{\rm evap}}\right)^{1/4}=a_{\rm eq}\left(\frac{\rho_c/a_{\rm eq}^3}{3M_{\rm Pl}^2/t_{\rm evap}^2}\right)^{1/4}\simeq 7\times 10^{-32}\left(\frac{M_{\rm PBH}}{M_{\rm Pl}}\right)^{3/2}
\end{equation}
where we used Eq.~(\ref{eq:evap}) in the next-to-last equality. The present velocity of dark matter produced from the evaporation of a PBH of mass $M_{\rm PBH}$ is thus
\begin{equation}
    v_{\chi}=\frac{p_{\rm now}}{m_{\chi}}\simeq 4\times 10^{-31}\left(\frac{m_{\chi}}{M_{\rm Pl}}\right)^{-1}\left(\frac{M_{\rm PBH}}{M_{\rm Pl}}\right)^{1/2}.
\end{equation}
Assuming that only redshift contributes to setting the current dark matter velocity, using the constraint of Ref.~\cite{Viel:2005qj} on the velocity of thermal relics today, 
\begin{equation}\label{eq:betalim}
    v_{\chi}\lesssim 4.9\times 10^{-7},
\end{equation}
we have
\begin{equation}\label{eq:limtbeta}
    \frac{m_{\chi}}{1\ {\rm GeV}}\gtrsim 2\times 10^{-6}\left(\frac{M_{\rm PBH}}{M_{\rm Pl}}\right)^{1/2}.
\end{equation}

The constraint above, together with the minimal primordial black hole mass allowed by CMB results, Eq.~(\ref{eq:hubbleconst}),  sets the minimal possible dark matter mass, if produced from evaporation,
\begin{equation}
    m_{\chi}\gtrsim 0.6\ {\rm MeV}.
    \label{eq:DMlimitvelocity}
\end{equation}
Notice that the constraint in Eq.~(\ref{eq:DMlimitvelocity}) applies only if a substantial fraction of the dark matter is produced by evaporation from PBH's of mass $M_{\rm PBH}$. Assuming that such fraction is, say, 10\% of the dark matter in the universe, given that the maximal density of dark matter from evaporation corresponds to the dark matter dominating the number of degrees of freedom PBH evaporate to, our constraint applies to (for instance for $m_{\chi}<M_{\rm Pl}^2/M_{\rm PBH}$)
\begin{equation}
    \beta\gtrsim 0.1 \frac{\rho_c}{m_{\chi} s_{\rm now}}\left(\frac{g_*}{100}\right)^{1/4}\left(\frac{g_\chi}{g}\right)\left(\frac{M_{\rm PBH}}{M_{\rm Pl}}\right)^{-1/2}.
\end{equation}
Notice that this is likely a fairly conservative constraint, as the limit in Eq.~(\ref{eq:betalim}) assumes 100\% of the dark matter has the quoted velocity.

\subsection{Baryon Asymmetry from PBH evaporation}\label{sec:bau}
Here, we consider three classes of models for baryogenesis via PBH evaporation: baryogenesis via non-thermal leptogenesis,  baryogenesis via the decay of grand unification gauge bosons (GUT baryogenesis) and, finally, we entertain the possibility that the dark matter is produced in conjunction with an asymmetry in the baryon sector.

In all cases, we determine whether PBH evaporation can lead to  the observed baryon asymmetry, with a baryon-number-to-entropy density of \cite{Aghanim:2018eyx}
$$
n_B/s\approx 8.8\times 10^{-11}.
$$
\subsubsection{Baryogenesis via leptogenesis}\label{sec:leptobaryo}
In the case of baryogenesis via leptogenesis,
\begin{equation}
    \frac{n_B}{s}=N_\nu\ \varepsilon\ \kappa \ Y_i\simeq 0.65 N_\nu\ \varepsilon\ \kappa\ \beta  \left(\frac{g_*}{100}\right)^{-1/4}\left(\frac{M_{\rm Pl}}{M_{\rm PBH}}\right)^{3/2},
\end{equation}
with 
\begin{equation}\label{eq:Nnu}
    N_\nu\simeq\frac{g_\nu}{g}\int_{M_{\rm PBH}}^0\frac{-dM}{3T}=\frac{g_\nu}{g}\int_{T_0}^\infty\frac{M_{\rm Pl}^2}{3T^3}dT=\frac{g_\nu}{6g}\left(\frac{M_{\rm PBH}}{M_{\rm Pl}}\right)^2,\quad M_\nu<T_H=M_{\rm Pl}^2/M_{\rm PBH}
\end{equation}
\begin{equation}
    N_\nu\simeq\frac{g_\nu}{g}\int_{M_\nu}^\infty\frac{M_{\rm Pl}^2}{3T^3}dT=\frac{g_\nu}{6g}\left(\frac{M_\nu}{M_{\rm Pl}}\right)^{-2},\quad M_\nu>T_H,
\end{equation}
and where $M_\nu$ is the right-handed neutrino mass scale (for simplicity we assume all right-handed neutrinos to be close-to-degenerate in mass), with $\varepsilon$ the $CP$ asymmetry factor of the right-handed neutrino decays, and with $\kappa\approx 0.35$ the conversion ratio of leptons to baryons \cite{Turner:1979bt}. An important constraint for the baryogenesis-via-leptogenesis scenario is that the inverse-decay of right handed neutrinos be out of equilibrium. This is guaranteed if the temperature of the universe at PBH evaporation is smaller than $M_\nu$, i.e. if
\begin{equation}\label{eq:Mnulim}
  M_\nu>T_{\rm evap}\simeq  1.9\times 10^{18}\ {\rm GeV} \left(\frac{M_{\rm Pl}}{M_{\rm PBH}}\right)^{3/2},
  \end{equation}
  where we assumed $g_*\simeq g\simeq 100$.  
  In the baryogenesis-via-leptogenesis scenario we also require that the evaporation temperature be larger than the electroweak scale, under which sphaleron rates are highly suppressed, thus enforcing
  \begin{equation}\label{eq:sphalerons}
      T_{\rm evap}\gtrsim100\ {\rm GeV} \Rightarrow M_{\rm PBH}\lesssim 7.1\times 10^{10}\ M_{\rm Pl}\simeq 1.4\times 10^6\ {\rm grams}.
  \end{equation}

In a model-independent way, the parameter $\varepsilon$ is {\em a priori} unconstrained and of ${\mathcal O} (1)$. However, in specific model realizations, $\varepsilon$ can be bounded from above. For instance, for type I seesaw models, barring tuned right-handed neutrino Yukawa textures \cite{Flanz:1996fb, Pilaftsis:1997jf}, one has \cite{Buchmuller:2002rq}
  \begin{equation}
      \varepsilon<\frac{3M_\nu m_{\rm max}}{16\pi v^2}\simeq 240\left(\frac{M_\nu}{M_{\rm Pl}}\right)\left(\frac{m_{\rm max}}{0.05\ {\rm eV}}\right),
  \end{equation}
  with $v$ the electroweak vacuum expectation value, and $m_{\rm max}$ the mass of the heaviest left-handed neutrino. In light of that and of cosmological constraints on $m_{\rm max}$ \cite{Aghanim:2018eyx}, and in this specific model context, only for $M_\nu\simeq 10^{14}$ GeV could $\varepsilon\sim{\mathcal O}(1)$, but not for smaller right-handed neutrino masses. However, larger phases are generically possible, see e.g. fig.~4 of Ref.~\cite{Pilaftsis:1997jf}. In what follows we consider a model independent scenario, and in order to show the maximal possible range of viable parameters, we set here  $\varepsilon=0.5$.
  
  Requiring a baryon asymmetry yield matching observations, and assuming $M_\nu>T_H$ and $g_\nu=6$, we get the following relation between the right-handed neutrino mass scale, the PBH mass and $\beta$ 
  \begin{equation}
      \beta\simeq \frac{2.3\times 10^{-9}}{\varepsilon\ \kappa}\left(\frac{M_{\rm Pl}^7}{M_\nu^4M_{\rm PBH}^3}\right)^{-1/2}.
  \end{equation}
  \subsubsection{GUT baryogenesis}\label{sec:GUTbaryo}
In the scenario where baryogenesis originates from the $CP$ and $B$-number violating decays of a GUT boson $X$, carrying $g_X$ degrees of freedom, the produced baryon asymmetry depends on the $CP$ violating parameter \cite{Baumann:2007yr}
\begin{equation}
    \gamma\equiv \sum_i\ B_i\ \frac{\Gamma(X\to f_i)-\Gamma(\bar X\to \bar f_i)}{\Gamma_X},
\end{equation}
where $B_i$ is the baryon number of the particular final state $f_i$, and $\Gamma_X$ the $X$ decay width. The expression for the resulting baryon asymmetry is then simply
\begin{equation}
    \frac{n_B}{s}=N_X\ \gamma \ Y_i\simeq 0.65 N_X\ \gamma \beta\   \left(\frac{g_*}{100}\right)^{-1/4}\left(\frac{M_{\rm Pl}}{M_{\rm PBH}}\right)^{3/2},
\end{equation}
with, just as above,
\begin{eqnarray}
    &&N_X\simeq\frac{g_X}{6g}\left(\frac{M_{\rm PBH}}{M_{\rm Pl}}\right)^2,\quad M_X<T_H=M_{\rm Pl}^2/M_{\rm PBH}\\
 &&\quad\ \ \ \simeq\frac{g_X}{6g}\left(\frac{M_X}{M_{\rm Pl}}\right)^{-2},\quad M_X>T_H,
\end{eqnarray}
We consider a fairly generous range for the mass scale $M_X$ of the GUT gauge bosons $X$ whose decay is responsible for the generation of the baryon asymmetry, 
\begin{equation}
    10^{15}\lesssim M_X/{\rm GeV}\lesssim 10^{17};
\end{equation} a variety of mechanisms can shift the precise energy scale of gauge coupling unficiation, and even when that scale is fixed, $M_X$ is not exactly determined (see e.g. Ref.~\cite{Hall:2014vga} and references therein). In the plots, we use $g_X=25$ and $\gamma=0.1$.

Notice that GUT baryogenesis requires a source of $B-L$ violation to prevent sphaleron washout of the produced baryon asymmetry for models where evaporation happens before the electroweak phase transition, here for masses $M_{\rm PBH}\lesssim 10^6$ g. We postulate in this case the mechanism outlined in Ref.~\cite{Fukugita:2002hu}, which posits the existence of heavy right-handed neutrinos interacting with the Standard Model Higgs doublet via an effective dimension five operator; as long as the induced lepton-number violating reaction is fast compared to the sphaleron rate (which is generically the case at high enough temperatures) then the $\Delta L$ component of the generated lepton-baryon asymmetry is erased, leaving a net $\Delta B$ which is unaffected by sphaleron washout.

\subsubsection{Asymmetric Dark Matter}
Finally, we consider a simple incarnation of asymmetric dark matter, inspired by the scenario detailed in Ref.~\cite{Falkowski:2011xh}. Schematically, the Standard Model is augmented with a dark-sector scalar field $\phi$ and a Dirac fermion $\chi$ coupled to right-handed neutrinos $N_i$, with  Lagrangian density
\begin{equation}
    -{\mathcal L}=-{\mathcal L}_{\rm SM}+\frac{1}{2}M_iN_i^2+Y_{i\alpha}N_iL_\alpha H+\lambda_iN_i\chi\phi+{\rm h.c.}
\end{equation}
plus mass terms for the $\phi$ and $\chi$. $\chi$ has lepton number +1, and $\chi$ and $\phi$ are charged under a discrete $Z_2$ symmetry that ensures the stability of the lightest dark sector state; we assume $m_{\chi}<m_\phi$, so $\chi$ is the stable species\footnote{Note that lepton number conservation forces $\chi$ to be a Dirac fermion, and to get mass from another fermion $\tilde\chi$ with opposite lepton number.}. We also need to assume fast, lepton-conserving interactions that thermalize leptons $l$, the Higgs, and the dark sector fields, annihilating away the symmetric components $l+\bar l$ and $\chi+\bar\chi$ (including, here, those non-thermally produced by PBH evaporation). Since the symmetric component of the dark matter must annihilate away by hypothesis in this scenario, the dark matter will generically reach kinetic equilibrium , thus reducing its velocity. As a result, the limit in Eq.~(\ref{eq:limtbeta}) does not apply here.

The $N_i$ decays are CP-violating, and the resulting decay asymmetries are defined summing upon Standard Model generations $\alpha=1...3$, $\epsilon_L=\sum_\alpha\epsilon_{L_\alpha}$, where
\begin{equation}
    \epsilon_\chi=\sum_i\frac{\Gamma(N_i\to\chi\phi)-\Gamma(N_i\to\bar\chi\phi^\dagger)}{\Gamma_{N_i}},\quad \epsilon_L=\sum_i\frac{\Gamma(N_i\to l h)-\Gamma(N_i\to\bar lh^\dagger)}{\Gamma_{N_i}}.
\end{equation}
The final asymmetry in each sector does not only depend on the decay asymmetries above, but also by  on the details of the models and on washout and transfer effects, which, following Ref.~\cite{Falkowski:2011xh}, we parameterize with the quantities $\eta_L$ and $\eta_\chi$ in the two sectors, respectively. Finally, the asymptotic asymmetries must satisfy \cite{Harvey:1990qw}
\begin{eqnarray}
    \label{eq:YDL}Y^\infty_{\Delta L}&=&\epsilon_L\eta_L N_\nu Y_i=\left(\frac{n_B}{s}\right)\frac{37}{12}\simeq 2.7\times 10^{-10}\\
    \label{eq:YDchi}Y^\infty_{\Delta\chi}&=&\epsilon_\chi\eta_\chi N_\nu Y_i\simeq 4.4\times 10^{-10}\left(\frac{1\ \rm GeV}{m_{\chi}}\right),
    \label{eq:admabundance}
\end{eqnarray}
where $Y_i$ is the same as what given in Eq.~(\ref{eq:yi}), $N_\nu$ is as given in Eq.~(\ref{eq:Nnu}) and, again as above, we assume the $N_i$ to be out of equilibrium and produced from PBH evaporation (thus with a mass satisfying the constraints of Eq.~(\ref{eq:Mnulim})).

In the case, for instance, where $M_\nu<T_H$, and thus $N_\nu$ is independent of $M_\nu$, we find that 
\begin{equation}\label{eq:NnuYi}
    N_\nu Y_i\simeq 0.04\beta\left(\frac{M_{\rm PBH}}{M_{\rm Pl}}\right)^{1/2}    
\end{equation}
and thus, given a value for $\epsilon_L\eta_L$, there is one value of $\beta$ that satisfies Eq.~(\ref{eq:YDL}), namely
\begin{equation}
    \beta\simeq\frac{6.8\times 10^{-9}}{\epsilon_L\eta_L}\left(\frac{M_{\rm PBH}}{M_{\rm Pl}}\right)^{-1/2}.   
\end{equation}
In turn, given $N_\nu Y_i$ as in Eq.~(\ref{eq:NnuYi}), there is a one-to-one correspondence between $\epsilon_\chi\eta_\chi$ and $m_{\chi}$ via Eq.~(\ref{eq:YDchi}). Specifically,
\begin{equation}
    \frac{\epsilon_L\eta_L}{\epsilon_\chi\eta_\chi}\equiv r_{L\chi}\simeq 0.61 \left(\frac{m_{\chi}}{1\ \rm GeV}\right).
\end{equation}

\section{Results}\label{sec:results}
We discuss in this section all of our numerical results for the framework described above. Sec.~\ref{sec:res_f0} assumes complete PBH evaporation and no Planck-scale relics (thus, $f=0$, where $f$ indicates the mass of PBH relics from evaporation in units of the reduced Planck mass); we show results for both the baryogenesis via leptogenesis (see sec.~\ref{sec:leptobaryo}) and for the GUT baryogenesis (sec.~\ref{sec:GUTbaryo}) scenarios, for a variety of dark matter masses; the following sec.~\ref{sec:res_f} assumes $f\neq 0$, and thus the existence of Planck-scale relics contributing to the global cosmological dark matter density, again for both baryogenesis scenarios, and again for a variety of dark matter masses; finally, in sec.~\ref{sec:ADM} we show results for asymmetric dark matter, for two different values of the right-handed neutrino mass scale. 

\subsection{Baryogenesis and Dark Matter from (complete) PBH evaporation}\label{sec:res_f0}
As outlined above, we assume exclusive non-thermal dark matter production from PBH evaporation, and we also assume that the dark matter never thermalizes. We intend to address two questions:\\[-0.3cm]

(1) What is the range of viable dark matter masses?\\[-0.3cm]

(2) Can dark matter and baryogenesis both be accounted for from PBH evaporation? If so, for which PBH masses?\\[-0.3cm]

We outlined above general constraints on the dark matter mass: the lower limits stems from Eq.~(\ref{eq:DMlimitvelocity}), while the upper limit corresponds to the maximal mass that can be produced from the evaporation of a PBH of mass $M_{\rm PBH}$; the upper limit lies in the regime where $M_X>T_H=M_{\rm Pl}^2/M_{\rm PBH}$ (for $M_X<T_H$, $M_X<M_{\rm Pl}/(9.1\times 10^4)$, because of Eq.~(\ref{eq:mpbh})), and is given by the requirement that $N_X>1$; The maximal possible $N_X$ corresponds to $g_X, g\to\infty$ and thus to $M_X<M_{\rm Pl}/\sqrt{6}\simeq 10^{18}\ {\rm GeV}$.

\begin{figure}[]\centering
\includegraphics[width=0.9\textwidth]{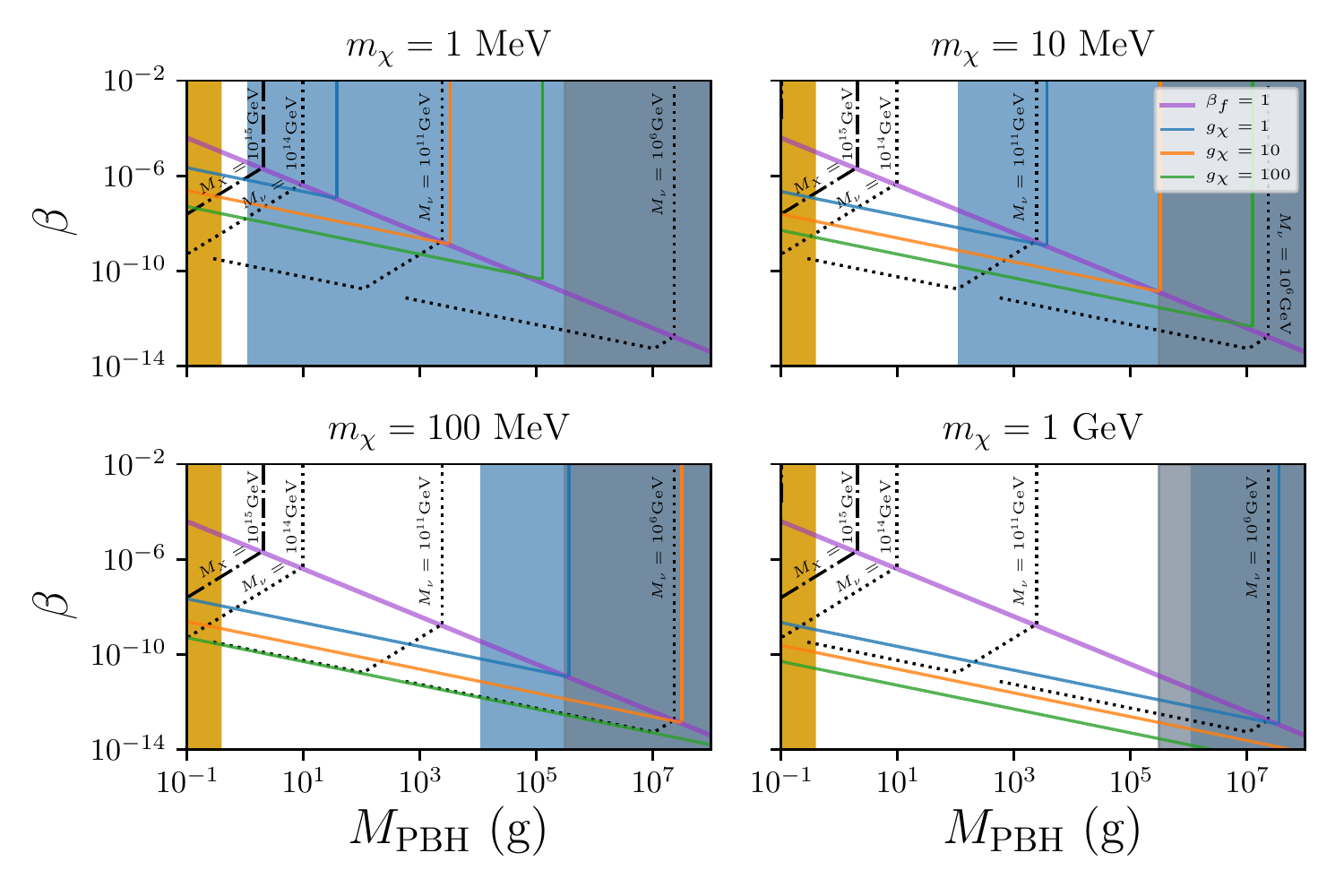}
\includegraphics[width=0.9\textwidth]{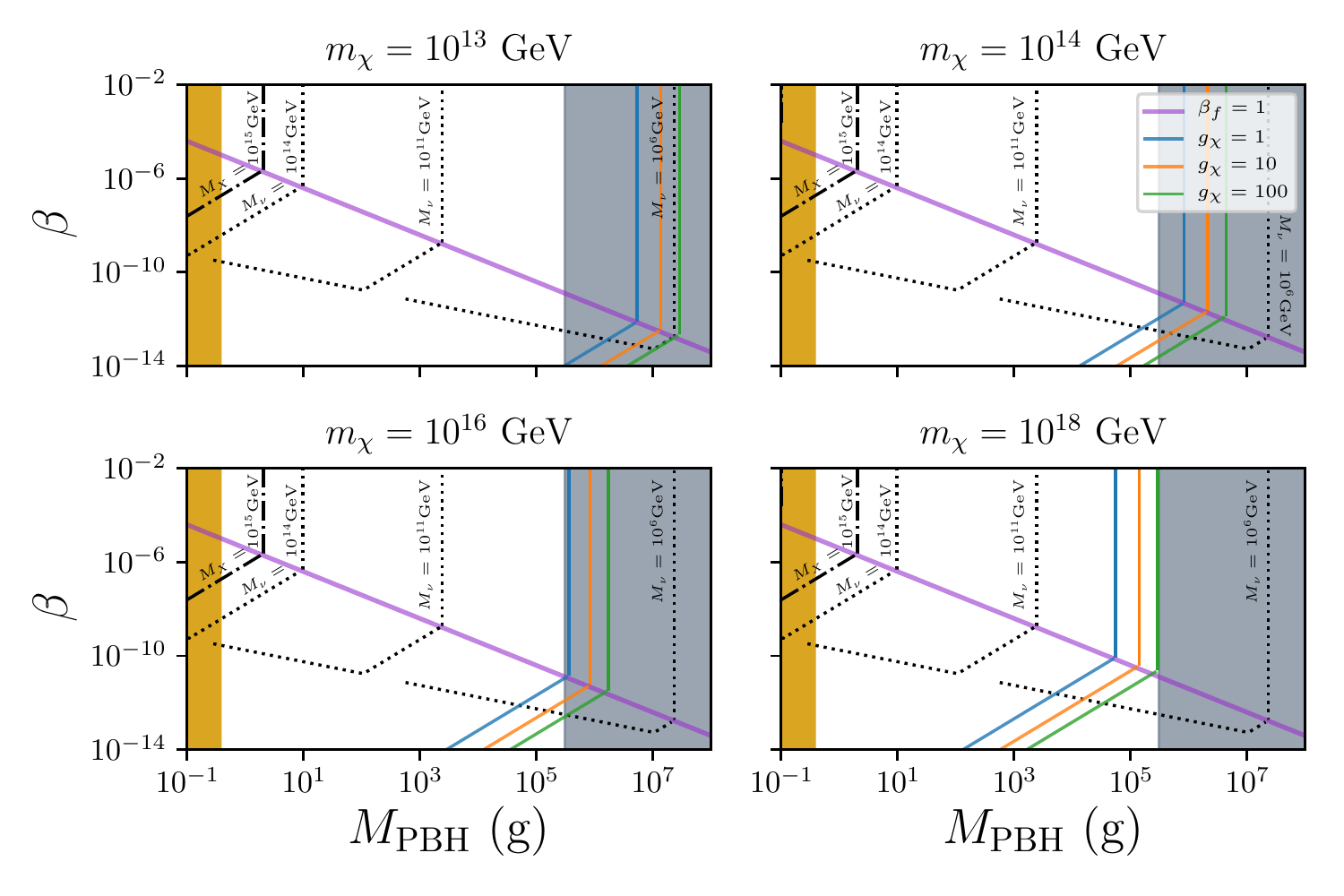}
\caption{\footnotesize Regions of successful production of the observed baryon asymmetry and dark matter on the ($M_{\rm PBH},\beta)$ plane. The region shaded in yellow on the left is ruled out by the CMB constraint of Eq.~(\ref{eq:hubbleconst}). The blue-shaded region is ruled out by the constraint on the velocity of the dark matter at late times; finally, the grey region violates the constraint of Eq.~(\ref{eq:sphalerons}), relevant for the leptogenesis scenario. The thick purple line corresponds to $\beta_f=1$ for $g_\chi=1$ plus Standard Model degrees of freedom, i.e. for $\beta$ above that line, PBH eventually dominate the energy density of the universe prior to evaporation. The colored lines correspond to the dark matter mass indicated on top of each panel and varying number of dark-sector degrees of freedom, as indicated in the legend. The dot-dashed lines indicate regions of successful baryogenesis via GUT bosons decay. Finally, the dotted lines indicate regions of successful baryogenesis via leptogenesis, corresponding to different right-handed neutrino mass scales, as indicated, and to a large CP violation parameter  $\varepsilon=0.5$ (smaller CP parameters would shift the curves to proportionally larger values of $\beta\sim 1/\varepsilon$).}
\label{fig:MeV}
\end{figure}

We present our results in Fig.~\ref{fig:MeV}. All plots in our study utilize the same parameter space: the ($M_{\rm PBH},\beta)$ plane (as a reminder, $\beta$ is the relative energy density of primordial black holes at the time of their genesis). In the plots, we shade in yellow the region at low PBH masses ruled out by the CMB limit of Eq.~(\ref{eq:mpbh}) from the lowest possible Hubble rate during inflation; we shade in blue the region excluded by the current dark matter velocity limit, Eq.~(\ref{eq:DMlimitvelocity}); finally, we shade in grey the region where PBH evaporation ends below the electroweak scale, and thus baryogenesis via leptogenesis is not effective because of suppressed sphaleron rates, Eq.~(\ref{eq:sphalerons}).

We indicate with a thick purple line the region where $\beta_f=1$ for $g_\chi=1$ plus Standard Model degrees of freedom, i.e. for $\beta$ above that line, PBH eventually dominate the energy density of the universe prior to evaporation, and particle production becomes $\beta$-independent (hence the lines corresponding to successful baryogenesis and dark matter production become vertical).

In the plots, the colorful solid lines correspond to different numbers of dark-sector degrees of freedom: the upper blue line corresponds to $g_\chi=1$, the orange line to $10$ and the green to 100. The dot-dashed line shows the parameter space compatible with GUT baryogenesis for $M_X=10^{15}$ GeV. Finally, the dotted lines correspond to baryogenesis via leptogenesis with right-handed neutrino mass scales $M_\nu=10^{14}$ GeV (upper line), $M_\nu=10^{11}$ GeV (middle line) and $M_\nu=10^{6}$ GeV (lower line). We truncate the dotted lines in this plot and in the following plots at PBH masses such that the corresponding non-thermally produced neutrinos would thermalize, thus violating Eq.~(\ref{eq:Mnulim}). Notice that for intermediate values of $M_\nu$, the envelope giving the lowest possible $\beta$ for successful leptogenesis is uninterrupted, and that values of $M_\nu<10^6$ GeV are also possible. Once again, viable leptogenesis occurs in the region encompassed by the dotted lines. 

We start with a dark matter mass of 1 MeV in the upper left panel of the top-four plots in figure \ref{fig:MeV}. This mass is only slightly above the limit in Eq.~(\ref{eq:DMlimitvelocity}) (incidentally, we note that slightly lower masses, between 0.6 and 1 MeV, are possible, but correspond to very narrow viable parameter space in $M_{\rm PBH}$). For such light masses, the constraint on the present dark matter velocity, from Eq.~(\ref{eq:betalim}), is quite stringent, and pushes against the constraint on $M_{\rm PBH}$ in Eq.~(\ref{eq:mpbh}). Notice that despite the fact that lighter black holes have a larger temperature, the earlier evaporation time means the produced dark matter has more time to cool by redshifting.

Light dark matter particles means, via Eq.~(\ref{eq:dmabundance}), that larger values of $\beta$ are needed at a given $M_{\rm PBH}$. In turn, this makes it easier to combine the generation of dark matter from evaporation and of the observed baryon asymmetry. Fig.~\ref{fig:MeV} shows that for dark matter masses at around 1 MeV, both leptogenesis and GUT baryogenesis work, with the former suitable for a large number of degrees of freedom, and the latter for a low number of degrees of freedom. PBH dominance of the universe's energy density forces $g_\chi>1$ in this case. The PBH mass needs to be right around 1 gram for a dark matter of 1 MeV. For sub-MeV dark matter masses we find that the only viable scenario is GUT baryogenesis, for PBH mass slightly below 1 gram, and again $g_\chi\sim 10$.

The upper right plot shows $m_{\rm DM}\sim 10$ MeV. GUT baryogenesis is now no longer possible, while there is substantial overlap with leptogenesis across a fairly extended range of PBH masses from 0.5 to around 100 grams. The same applies to 100 MeV dark matter masses, although here the viable PBH mass range is extended to larger values, up to the limit from the current dark matter velocity, which, for a dark matter mass of 100 MeV, is around a few tens of kg. Finally, for DM masses at the GeV (lower right panel) or more, the parameter space keeps enlarging as the constraint on the dark matter velocity weakens (for 1 GeV up to ton-scale PBH); leptogenesis remains viable as long as the right-handed neutrino mass is sufficiently low.

In the lower four panels, we show a different regime, where the dark matter mass is very heavy ($10^{13}$ to $10^{18}$ GeV), and is produced by PBH whose initial temperature is {\em lower} than the dark matter mass. In this regime, the dependence with $M_{\rm PBH}$ is no longer $\Omega_\chi\sim M_{\rm PBH}^{1/2}$ but is instead $\Omega_\chi\sim M_{\rm PBH}^{-3/2}$ (see Eq.(\ref{eq:dmabundance2})). The top two panels illustrate that the dark matter mass must be at least a few $\times 10^{13}$ to be viable, with generally very low right-handed neutrino masses. Notice that the constraint on the PBH mass from evaporation ending prior to the EW phase transition, forces the lowest dark matter mass to be heavier than a few times $10^{12}$ GeV. Also, notice that GUT baryogenesis is {\em never} an option for very heavy dark matter masses.

\subsection{Baryogenesis and Dark Matter from PBH evaporation and PBH relics}\label{sec:res_f}
Here we discuss the possibility that evaporation stops at a mass $f M_{\rm Pl}$, leaving the dark matter produced by the PBH evaporation {\em together} with a second population of stable Planck-scale relics of mass $M_{\rm relic}=f\times M_{\rm Pl}$; we explore this two-component dark matter scenario on the same parameter space as before, taking into consideration the over-closure constraint from the PBH relics (the corresponding excluded region of parameter space is shaded in dark red, and is at the top left of the plots).

\begin{figure}[!h]\centering
\includegraphics[width=0.9\textwidth]{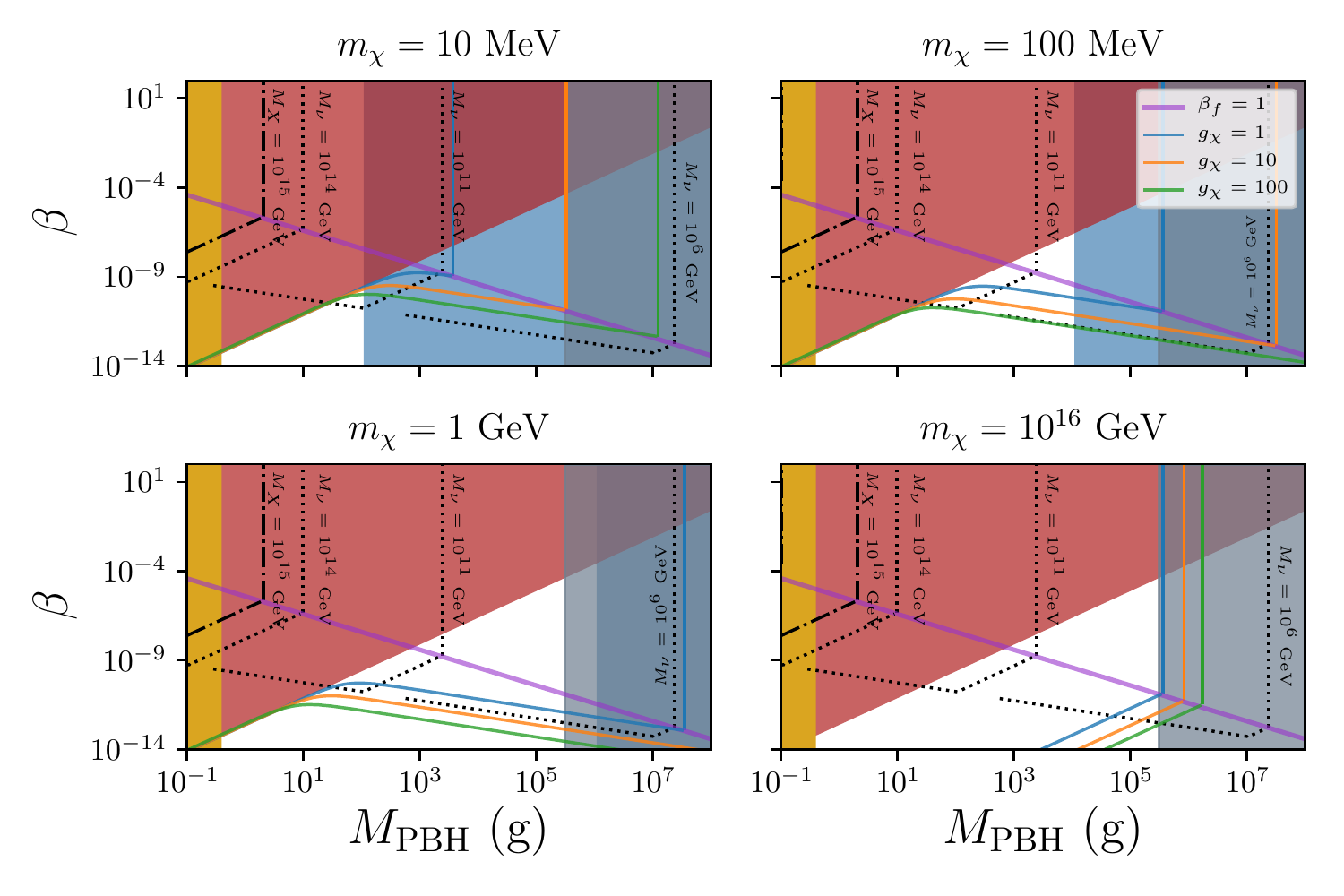}
\includegraphics[width=0.9\textwidth]{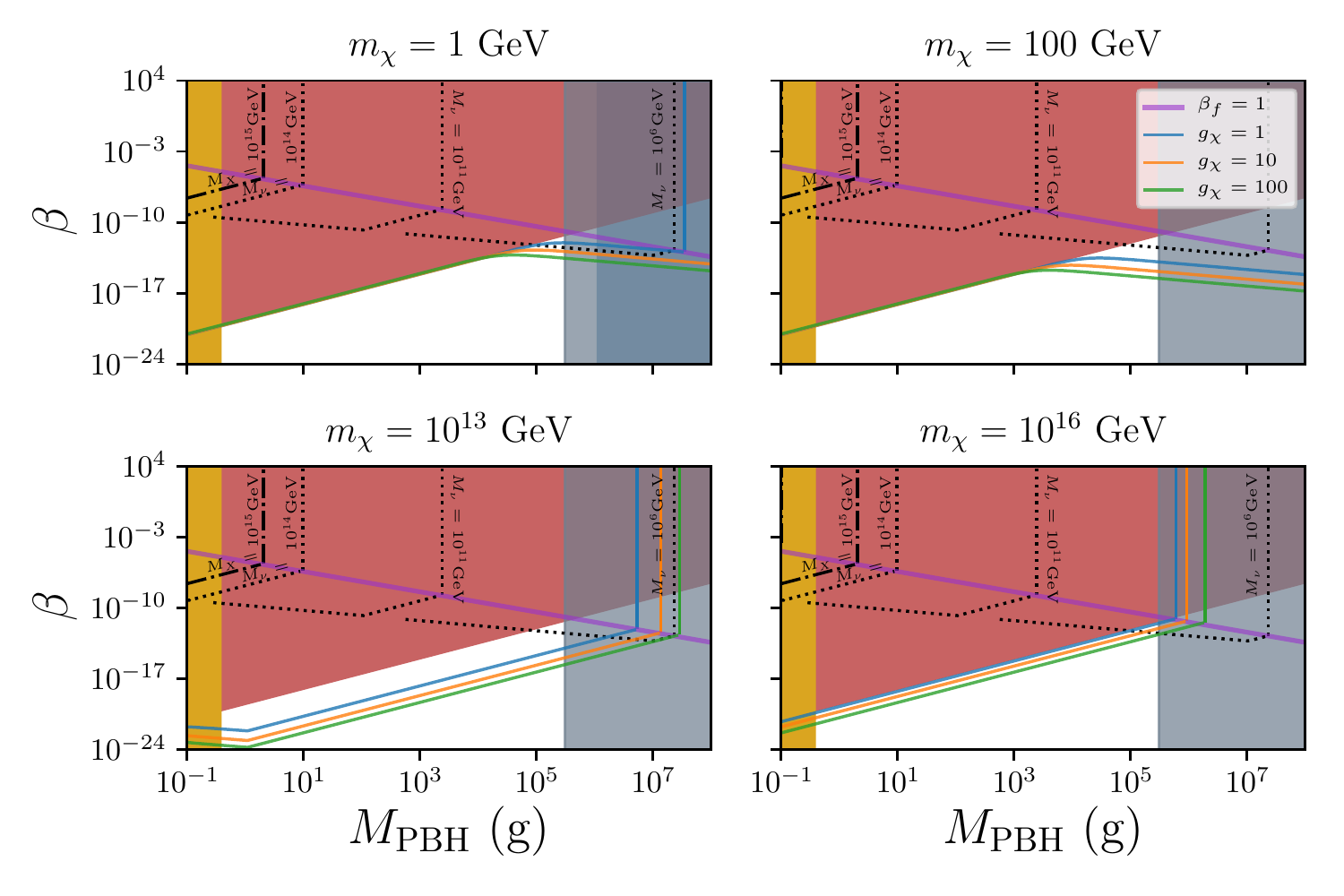}
\caption{Top: Mixed-dark matter case, with Planck-scale relic of mass $M=fM_{\rm Pl}$, and $f=10^{-7}$. The shaded region in the upper left indicates an excessive density of Planck-scale relics, all other lines are the same as in fig.~\ref{fig:MeV}. Bottom: same, for $f=1$ (notice the different $y$ axis).}
\label{fig:f}
\end{figure}

In the top four panels of fig.~\ref{fig:f} we show the case where $f=10^{-7}$. This is, admittedly, a very low mass scale for PBH evaporation relics, but given ignorance about how evaporation might stop due to quantum gravity effects, it cannot be {\em a priori} ruled out. For such light PBH relics, we find that successful baryogenesis (via leptogenesis) plus two-component dark matter is possible for masses between roughly 10 MeV and a few GeV in the regime where $m_{\chi}<T_H$, and is possible again for very heavy dark matter $m_{\chi}>10^{13}$ GeV (in the figure, we show $m_\chi=10^{14}$ GeV), but in this case the contribution of Planck relics is very sub-dominant. The right-handed neutrino mass needs to be between $10^{7}$ and $10^{12}$ GeV, and the PBH mass between 1 g and around a ton for this scenario to be successful in the low-dark matter mass regime; the heavy dark matter, as before, demand low right-handed neutrino masses, around $10^6$ GeV or so. For dark matter larger than or around 10 GeV but lighter than around $10^{13}$ GeV, dark matter and the baryon asymmetry cannot be jointly produced; also, we find that GUT baryogenesis never works if Planck relics are around (the corresponding region of parameter space is ruled out by overclosure from the density of Planck relics, unless $f\to0$).

The lower four panels show the case where evaporation stops {\em at the Planck scale}, i.e. $f=1$. In this case the two-component dark matter is viable for a broad range of masses; demanding successful baryogenesis via leptogenesis forces the dark matter mass to be at the GeV scale (top left panel) and right handed neutrinos to be around 10 TeV; lighter dark matter particles in the MeV range are ruled out for $f\sim1$, as are heavier masses (see top left panel showing $m_{\chi}=100$ GeV). The bottom, left panel, with $m_{\chi}=10^{13}$ GeV, shows (at around $M_{\rm PBH}\sim 1$ g) the turnover of the regime when $m_{\chi}\sim T_H$; we do not find, however, regions of successful baryogenesis and dark matter; slightly heavier dark matter masses again make it possible to have successful leptogenesis, for sufficiently low right-handed neutrino masses (see bottom right panel, with $m_{\chi}=10^{16}$ GeV).

\begin{figure}[!h]\centering
\includegraphics[width=0.9\textwidth]{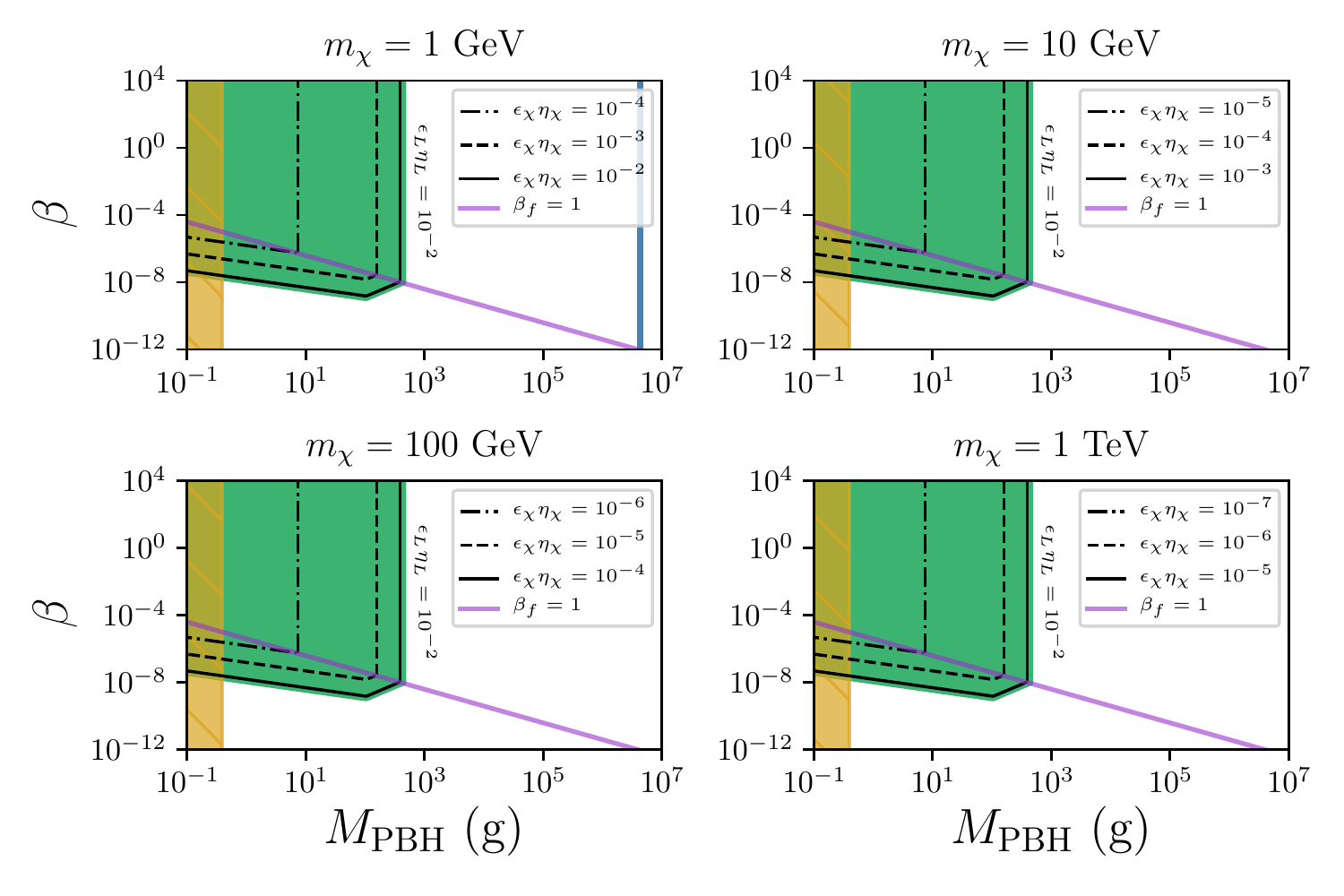}
\includegraphics[width=0.9\textwidth]{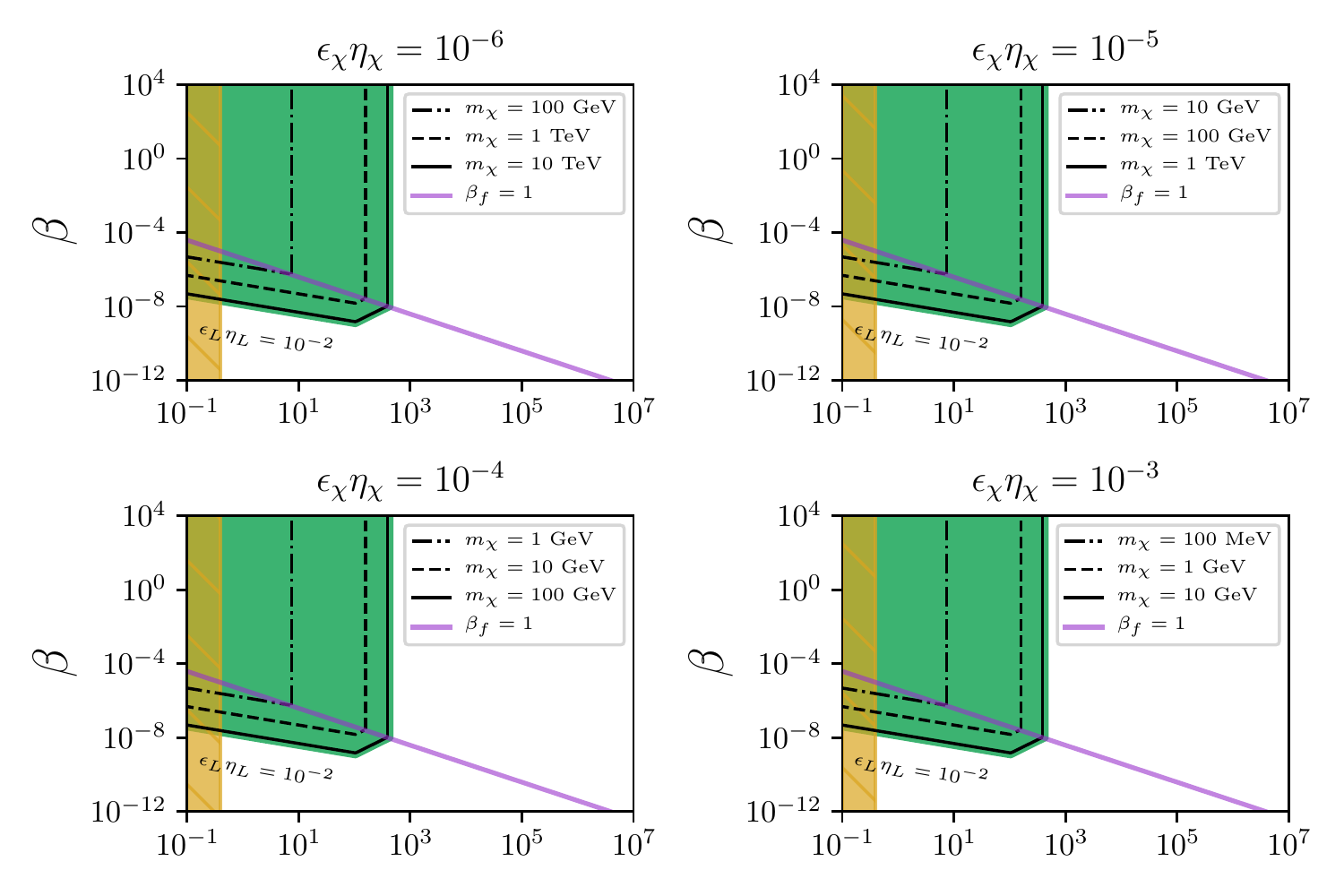}
\caption{The Asymmetric Dark matter scenario, with $M_\nu=10^{11}$ GeV. The green shaded region allows for successful asymmetric baryogenesis-via-leptogenesis, for $10^{-8}<\epsilon_L\eta_L<10^{-2}$. Each panel in the top four plots assumes a different dark matter, mass, $m_{\chi}=1$ GeV, 10 GeV, 100 GeV and 1 TeV. The black lines in those plots show the required values of $\epsilon_\chi\eta_\chi$ to produce the observed density of (asymmetric) dark matter. In the lower four panels, we instead fix $\epsilon_\chi\eta_\chi$ to several different values, $10^{-6},\ 10^{-5},\ 10^{-4},\ 10^{-3}$, and show lines corresponding to values of the dark matter mass that, in turn, would produce the observed dark matter density. As before, the purple line indicates $\beta_f=1$. The vertical dark blue line shows the limit from the dark matter velocity calculated as in Appendix \ref{app:a}.}
\label{fig:adm1}
\end{figure}

\begin{figure}[]\centering
\includegraphics[width=0.9\textwidth]{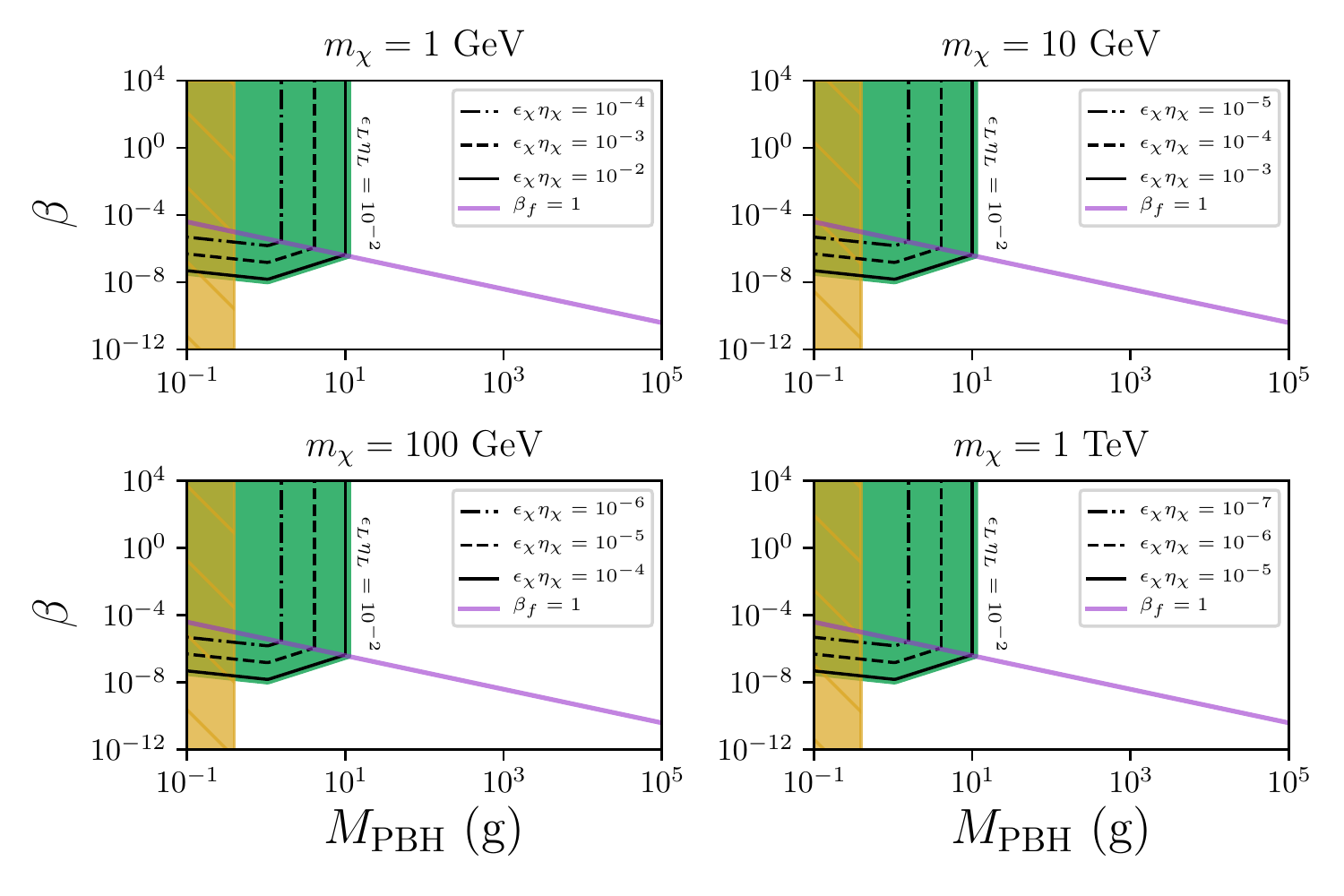}
\includegraphics[width=0.9\textwidth]{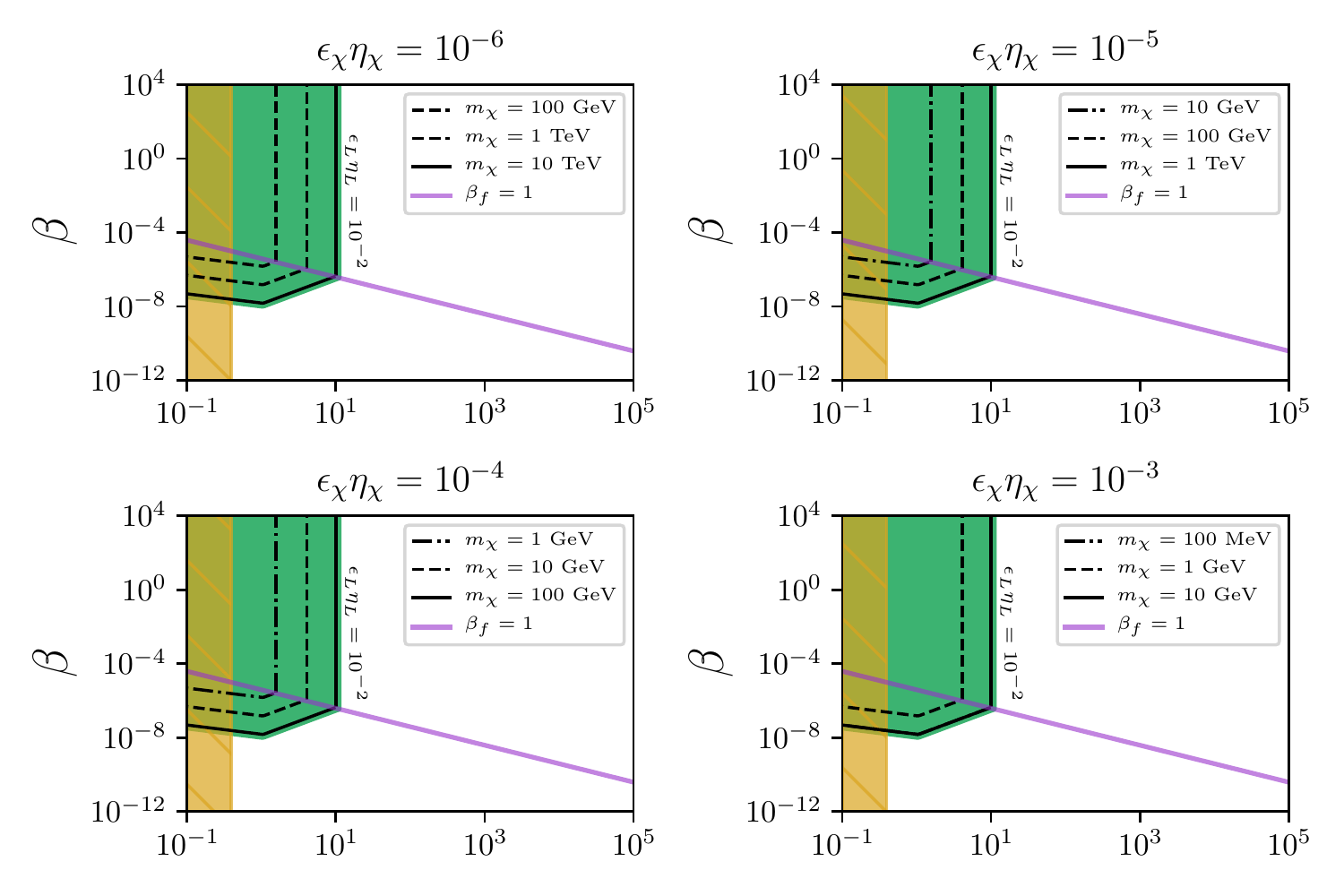}
\caption{As in fig.~\ref{fig:adm1}, but with $M_\nu=10^{13}$ GeV}
\label{fig:adm2}
\end{figure}

\subsection{Asymmetric Baryogenesis and Dark Matter from PBH evaporation}\label{sec:ADM}

In the asymmetric dark matter scenario, in addition to the plots' parameter space, i.e. the ($M_{\rm PBH},\beta)$ plane, the framework we consider has four additional parameters: the $CP$-asymmetry-washout-factor products $\epsilon_\chi\eta_\chi$ and $\epsilon_L\eta_L$, the dark matter mass $m_{\chi}$, and the right-handed neutrino mass $M_\nu$. We consider two representative right-handed neutrino mass scales, $M_\nu=10^{11}$ GeV  in fig.~\ref{fig:adm1} and $M_\nu=10^{13}$ GeV in fig.~\ref{fig:adm2}. In each of the top-four panels of each figure we fix the dark matter mass $m_{\chi}$ and show, on the ($M_{\rm PBH},\beta)$ plane the necessary values for $\epsilon_\chi\eta_\chi$ to reproduce the universe's observed dark matter abundance, superimposed with regions where the baryon asymmetry can be produced for a given range of $\epsilon_L\eta_L$. Specifically, for definiteness we shade in light green the region corresponding to 
\begin{equation}
    10^{-8}<\epsilon_L\eta_L<10^{-2}.
\end{equation}
(Notice that a broader range is theoretically possible). For a given dark matter mass, we find that there is ample parameter space to produce the observed dark matter density via PBH evaporation and subsequent asymmetric right-handed neutrino decay. Of course, from Eq.~(\ref{eq:admabundance}) it follows that the lower the dark matter mass, the larger the needed $\epsilon_\chi\eta_\chi$.

In fig.~\ref{fig:adm1}, the right-handed neutrino mass scale is low enough that for the relevant PBH mass range both $T_H>M_\nu$ and  $T_H<M_\nu$ are possible (the latter at masses larger than around 100 grams, the former for lighter masses), hence the shape of the green-shaded regions. The figure illustrates that a broad range of dark matter masses are possible, depending on model details fixing the $\epsilon_\chi\eta_\chi$ and $\epsilon_L\eta_L$ products.

In the bottom four panels, we fix the product $\epsilon_\chi\eta_\chi$ to several different values, namely $10^{-6},\ 10^{-5},\ 10^{-4},\ 10^{-3}$, and show lines corresponding to values of the dark matter mass that, in turn, would produce the observed dark matter density. Again, a wide range of values for the dark matter mass is possible, depending on the value of the parameter $\epsilon_\chi\eta_\chi$. The lower $\epsilon_\chi\eta_\chi$, the heavier the possible range of masses where the asymmetric dark matter and baryon asymmetry generation is possible.

We note that the limits on the current dark matter velocity are here different than before. First, the dark matter originates from right-handed neutrino decays rather than directly from evaporation. The right-handed neutrino lifetime is always much shorter than the PBH evaporation time scale; hence, effectively, the right-handed neutrino has no time to redshift, and the dark matter is produced by immediate subsequent decay. Assuming for simplicity isotropic decays in the rest frame of the neutrino, as we show in the Appendix the average dark matter velocity in this case is a factor 2 smaller than in the case of direct production from evaporation. As a result, the constraints on the dark matter mass are generically a factor 2 weaker (see Appendix~\ref{app:a}).

However, since in the asymmetric dark matter production scenario we posit that processes exist that deplete the symmetric $\bar\chi \chi$ component produced by evaporation, the dark matter velocity might, and will, be affected by such processes. For instance, should the depletion of the symmetric component proceed via annihilation with the visible sector, i.e. $\bar\chi\chi\to{\rm SM}$, then the dark matter would be presumably brought in kinetic equilibrium and thus {\em cool} to the visible sector temperature, weakening the limit discussed above; if, on the other hand, the depletion occurs via $2n\to2$ ``cannibal'' processes, with $n>1$, such as $\bar\chi\chi\bar\chi\chi\to\bar\chi\chi$,  then effectively the dark matter would {\em heat} itself up, strengthening the limits discussed above. 

For reference, in the figures we leave a vertical thick blue line in the top-left panel, corresponding to the heating from sterile neutrino decay, with the understanding that such limits are model-dependent and could be stronger or weaker than what the lines indicates. In practice, however, these constraints are largely outside the region of parameter space of interest.

Notice that a similar discussion to what we treat in the Appendix would be in order if the dark sector particles contained particles with masses largely different from the dark matter mass they eventually decay into. As mentioned above, here we make the simplifying assumption that the dark sector spectrum is trivially degenerate at a mass scale close to $m_{\chi}$.

\section{Discussion and Conclusions}\label{sec:conclusions}

We studied the joint production of the observed matter-antimatter asymmetry and of the cosmological dark matter from the evaporation of light primordial black holes (PBH) in the very early universe, at times $t_{\rm evap}\ll 1$ sec. The parameters of the model we considered include a universal mass for the primordial black holes, and their relative abundance at generation. We assumed that the dark matter belongs to a ``dark sector'' with a certain number of dark degrees of freedom. We also considered a ``mixed dark matter scenario'', where the dark matter is both produced by PBH evaporation and consists of PBH relics from the end of evaporation at around, or below, the Planck scale. Finally, we considered three scenarios for the generation of the matter-antimatter asymmetry: (i) $CP$- and $B$-violating decays of GUT gauge bosons, (ii) baryogenesis through (non-thermal) leptogenesis via out-of-equilibrium $CP$- and $L$- violating decays of heavy right-handed neutrinos, and (iii) asymmetric dark matter and baryogenesis, again via decays of heavy right handed neutrinos.

The parameter space under consideration is constrained by a variety of considerations, from limits on the dark matter velocity inherited from the large Hawking-Gibbons temperature scales at which the dark matter was produced, to limits on the PBH mass from the Hubble rate during inflation, to an excessive density of relic PBH from the end of evaporation at the Planck scale.

Unlike in previous studies that focused on scenarios where PBH dominate the energy density of the universe at production (see e.g. Ref.~\cite{Baumann:2007yr, Fujita:2014hha}), here PBH can be a subdominant component to the early universe's energy density, with generally different conclusions (although our results correspond to those of Ref.~\cite{Fujita:2014hha} for values of $\beta$ such that prior to evaporation PBH dominate the energy density of the universe).

If evaporation does not stop, and PBH vanish completely, both GUT baryogenesis and leptogenesis can be successful in conjunction with dark matter production from evaporation. GUT baryogenesis only works if the dark matter is between 1 and 10 MeV, while leptogenesis works either for dark matter masses between 1 MeV and a few GeV, or for super-heavy masses from $10^{13}$ to $10^{18}$ GeV, the maximal possible dark matter in this scenario. The needed PBH masses range from a few grams (for light dark matter particle masses), to around a ton for super-heavy dark matter.

If PBH evaporation does stop at some scale $f M_{\rm Pl}$, GUT baryogenesis is ruled out entirely, and leptogenesis works only for masses up to a few GeV or, again, for very heavy dark matter masses. In the former case, right handed neutrino masses must be large ($10^{11}$ GeV or so), in the latter, they must be much lighter ($10^6$ GeV or less).

Asymmetric dark matter and baryogenesis is successful, in this framework, for a broad range of effective $CP$ times washout factors $\epsilon\eta$ for the visible and dark sector. The larger the dark sector value of the product $\epsilon_\chi\eta_\chi$, the lighter the viable range of dark matter masses. The dark matter, in this scenario, must have a mass between a fraction of a GeV and 10 TeV or so.

If the scenario discussed here is indeed the backdrop for the generation of visible and dark matter, the detection outlook is relatively daunting. Searches for relic Planck-scale objects are possible, and in some cases might set some limits on this scenario, especially if the relic PBH are a substantial fraction of the dark matter, and/or if the relic are charged \cite{Drobyshevski:2012qk}. Directly or indirectly detecting the dark matter produced in PBH evaporation in the present scenario is problematic: since we assume no thermal equilibrium at any temperature, the indirect detection rates generically are highly suppressed, and so are the direct detection rates. 

One possible route to test this scenario (and in fact any scenario involving light PBH) is to look for gravitational wave emission from evaporation \cite{Dolgov:2011cq}: while all evaporation products quickly thermalize in our scenario, gravitons do not, leaving an imprint that is in principle detectable. There is a one-to-one correspondence between the frequency $\nu$ of gravity waves at the present time and the corresponding  frequency at emission $\nu_*$, emission which assume here to happen at the PBH evaporation time:
\begin{equation}\label{eq:freq}
    \nu\simeq0.34\ \nu_*\ \frac{T_0}{T_*}\left(\frac{100}{g_*(T_{\rm evap})}\right)^{1/3},
\end{equation}
where the temperature of the PBH evaporation $T_{\rm evap}$ is given in  Eq.~(\ref{eq:tevap}) and $T_0$ corresponds to the CMB frequency, around 160.4 GHz. As a result, we have
\begin{equation}
    \nu\simeq7.1\times 10^{10}\ {\rm Hz}\left(\frac{\nu_*}{M_{\rm Pl}}\right)\left(\frac{100}{g_*(T_{\rm evap})}\right)^{1/12}\left(\frac{g}{100}\right)^{-1/2}\left(\frac{M_{\rm PBH}}{M_{\rm Pl}}\right)^{3/2}.
\end{equation}
The maximal value for $\nu_*$ is around a few times the Hawking-Gibbons temperature, $T_H=M_{\rm Pl}^2/M_{\rm PBH}$. Using Eq.~(\ref{eq:freq}), we get that the maximal frequency of gravity waves today is around $10^{16}$ Hz. Generally, the spectrum peaks at $\nu_*\sim 2.8T_H$ \cite{Dolgov:2011cq}, therefore producing a signal at frequencies much higher than current gravity wave detectors. As we explain below, detection is however possible through the inverse Gertsenshtein effect \cite{ref67,ref66}.

We estimate here the strain corresponding to the predicted gravity wave signal from PBH evaporation. Ref.~\cite{Dolgov:2011cq} calculates that the energy density of gravity waves from PBH evaporation integrated over frequencies, and accounting for our assumption that PBH do not dominate the energy density of the universe, but rather constitute a fraction $\beta$ of it at production, is approximately
\begin{equation}
    \Omega_{\rm GW}(t_{\rm evap})\simeq 0.006\left(\frac{g_G}{100}\right)^2\beta,
\end{equation}
with $g_G=2$ the number of graviton degrees of freedom. The equation above also assumes graviton production to happen instantaneously at the evaporation time (see also Ref.~\cite{Fujita:2014hha}). The red-shifted gravitational wave density today is
\begin{equation}
    \Omega_{\rm GW}h^2(t_{\rm now})\simeq 1.67\times 10^{-5}\left(\frac{100}{g_*(T_{\rm evap})}\right)^{1/3}\Omega_{\rm GW}(t_{\rm evap})\simeq 10^{-11}\left(\frac{g_G}{10^{-2}}\right)^2\beta.
\end{equation}
The corresponding strain is then
\begin{equation}
h\sim 10^{-36}\beta^{1/2}.
\end{equation}
The frequencies for the gravity wave emitted from PBH evaporation are well beyond currently operating or future interferometers; however they could be detectable via the so-called inverse Gertsenshtein effect \cite{ref70}: The passage of gravity waves in a static magnetic field sources electromagnetic waves. As long as the induced signal ``beats'' thermal noise, a signal can be detected \cite{ref67, ref66}.

In conclusion, early evaporation of light primordial black holes can lead to co-genesis of a baryon asymmetry and of the dark matter. We demonstrated that several possible baryogenesis scenarios are viable (GUT baryogenesis, leptogenesis, asymmetric dark matter), for a broad range of dark matter masses and of primordial black hole masses. The dark matter itself can originate entirely from evaporation, or from decay of particles produced by PBH evaporation, or it can be a mix of particles from evaporation, and Planck-scale relics of the evaporation process. Detection prospects for the dark matter are discouraging, but this scenario would leave an imprint of very high-frequency gravitational waves, of calculable spectrum and intensity, possibly detectable via the inverse Gertsenshtein effect.

\section*{Acknowledgements}
We are grateful to Nicolas Fernandez, Benjamin V. Lehmann and John Tamanas for helpful comments; SP is grateful to his former undergraduate student Michael Shamma for pointing out a possible connection to asymmetric dark matter in this context. SP acknowledges gracious lessons in modern Greek neologisms from Vasiliki Pavlidou and John Tamanas. This work is partly supported by the U.S.\ Department of Energy grant number de-sc0010107.

\appendix
\section{Appendix: DM Velocity Constraint for Intermediate Decay through a Heavy Neutrino}\label{app:a}

Here we compute the average dark matter (DM) momentum from decay of a heavy neutrino \(N\) to a DM fermion \(\chi\) and a scalar \(\phi\). We will assume a flat matrix element for this process for simplicity. In the rest frame (RF) of the heavy neutrino, the four-momentum is given by
\begin{align}
	P^{\mu}_{N,RF} = (M_{N}, 0, 0, 0)
\end{align}
Consider the heavy neutrino in a frame boosted along the \(z\)-axis. Let's call this frame the lab-frame (LF). In this boosted frame, the four-momentum for the heavy neutrino is:
\begin{align}
	P^{\mu}_{N,LF} = (E_{N}, 0, 0, p_{N})
\end{align}
where 
\begin{align}
	E_{N} &= \gamma M_{N} & p_{N} &= \gamma v_{N} M_{N}
\end{align}
with \(\gamma\) and \(v_{N}\) being the relativistic boost factor and the velocity of the heavy neutrino, respectively. In the rest frame of the heavy neutrino, the DM four-momentum can be considered to be in the \(xz\)-plane, given by:
\begin{align}
	P^{\mu}_{\chi, RF} &= (E_{\chi}, p_{\chi}\sin\theta, 0, p_{\chi}\cos\theta)
\end{align}
In the LF, the DM four-momentum becomes
\begin{align}
	P^{\mu}_{\chi, LF} &= (\gamma E_{\chi} + v_{N}\gamma  p_{\chi}\cos\theta, p_{\chi}\sin\theta, 0, \gamma p_{\chi}\cos\theta + v_{N}\gamma E_{\chi})
\end{align}
In the case where both the DM and the scalar are ultra-relativistic, we have that 
\begin{align}
	E_{\chi} = p_{\chi} = M_{N} / 2
\end{align}
In this case, the four-momentum of the DM in the LF simplifies to:
\begin{align}
	P^{\mu}_{\chi, LF} &= \frac{M_{N}}{2}(\gamma + v_{N}\gamma  \cos\theta, \sin\theta, 0, \gamma\cos\theta + v_{N}\gamma)
\end{align}
The magnitude of the DM three-momentum is given by:
\begin{align}
	|\vec{P}_{\chi,LF}| = \frac{M_{N}}{2}\sqrt{\sin^{2}\theta + \gamma^{2}\left(\cos\theta + v_{N}\right)^{2}}
	= \gamma\frac{M_{N}}{2}\sqrt{(1-\beta^{2})\sin^{2}\theta + \left(\cos\theta + v_{N}\right)^{2}}
\end{align}
Expanding out the terms inside the square root, we find
\begin{align}
	(1-v_{N}^{2})\sin^{2}\theta + \left(\cos\theta + v_{N}\right)^{2} 
	&= \left(1+v_{N}\cos\theta\right)^{2}
\end{align}
where we used \(\gamma = \left(1-v_{N}^{2}\right)^{-1/2}\). Therefore:
\begin{align}
	|\vec{P}_{\chi,LF}|
	= \gamma\frac{M_{N}}{2}\left(1+v_{N}\cos\theta\right)
\end{align}
Averaging this expression over \(\theta\) from 0 to \(\pi\), we find 
\begin{align}
	\langle|\vec{P}_{\chi,LF}|\rangle = \dfrac{1}{\pi}\int_{0}^{\pi}|\vec{P}_{\chi,LF}|d\theta =  \gamma\frac{M_{N}}{2}
\end{align}
If the neutrino energy is given by \(6T_{H}\), then \(\gamma = 6T_{H}/M_{N}\) and 
\begin{align}
	\langle|\vec{P}_{\chi,LF}|\rangle = 3T_{H}
\end{align}
Combining this result with Eq.~(\ref{eqn:dm_mom_now}), Eq.~(\ref{eqn:aevap}), and  Eq.~(\ref{eq:betalim}), we find that 
\begin{equation}
    \dfrac{m_{\chi}}{1 \ \mathrm{GeV}} > 1\times10^{-6}\left(\dfrac{M_{\mathrm{PBH}}}{M_{\mathrm{Pl}}}\right)^{1/2}
\end{equation}

\bibliographystyle{JHEP}
\bibliography{references}

\end{document}